\documentclass[12pt, letterpaper]{article}
\usepackage[margin=1in]{geometry}
\pdfoutput=1
\usepackage{graphicx}
\usepackage{setspace}
\usepackage{biblatex}
\bibliography{arXiV.bib}

\begin{document}

\begin{center}
    {\large{\textbf{Detection of a space capsule entering earth's atmosphere with distributed acoustic sensing (DAS)}}}

Chris G. Carr, Carly M. Donahue\footnote{Corresponding author: cmd@lanl.gov}, Lo\"{i}c Viens, Luke B. Beardslee, Elisa A. McGhee, Lisa R. Danielson\\
CGC, CMD, LV, LBB, LRD: Los Alamos National Laboratory, Los Alamos, NM\\
EAM: Colorado State University, Fort Collins, CO 
\end{center}

\subsection*{Abstract} 
On 24 September 2023, the Origins, Spectral Interpretation, Resource Identification, and Security – Regolith Explorer (OSIRIS-REx) Sample Return Capsule entered the Earth's atmosphere after successfully collecting samples from an asteroid. The known trajectory and timing of this return provided a rare opportunity to strategically instrument sites to record geophysical signals produced by the capsule as it travelled at hypersonic speeds through the atmosphere. We deployed two optical-fiber distributed acoustic sensing (DAS) interrogators to sample over 12 km of surface-draped, fiber-optic cables along with six co-located seismometer-infrasound sensor pairs, spread across two sites near Eureka, NV. This campaign-style rapid deployment is the first reported recording of a sample return capsule entry with any distributed fiber optic sensing technology. The DAS interrogators recorded an impulsive arrival with an extended coda which had features that were similar to recordings from both the seismometers and infrasound sensors. While the signal-to-noise of the DAS data was lower than the seismic-infrasound data, the extremely dense spacing of fiber-optic sensors allowed for more phases to be clearly distinguished and the continuous transformation of the wavefront as it impacted the ground could be visualized. Unexpectedly, the DAS recordings contain less low-frequency content than is present in both the seismic and infrasound data. The deployment conditions strongly affected the recorded DAS data, in particular, we observed that fiber selection and placement exert strong controls on data quality. 

\subsection*{Introduction}

Approximately $10^5$ tons of extraterrestrial material enter the Earth’s atmosphere per year at hypersonic velocities ranging between 11.2 and 72.8 km/s \cite{Ceplecha1998, Plane2012}. While most particles are meteoroids with diameters less than one meter that disintegrate at altitudes above 20 km, meter-size and larger asteroids also enter the Earth's atmosphere and can produce shockwaves upon entering the upper regions of the atmosphere. Such objects can penetrate deep into the atmosphere and their fragments can even reach the Earth's surface as meteorites. A recent example is the Chelyabinsk superbolide, which deposited energy of approximately 500 kt of TNT equivalent, leaving a wake of destruction from its shockwave beneath its path and a $\sim$18$\,$m diameter crater \cite{Brown2013, Popova2013}. 

The prediction of shock waves and impacts from meter-size objects is difficult due to the spatio-temporal uncertainties related to their trajectories. Therefore, most recordings of objects entering the atmosphere using geophysical instruments on the ground (e.g., seismometers and infrasound sensors) are serendipitous and the instrument sparsity does not generally allow for a detailed characterization of their trajectory and recordings of the full wavefield. Space mission sample return capsules, which can re-enter from interplanetary space at speeds greater than 11 km/s, provide reasonable analogs for meter-sized natural meteors. The planned character of space capsule re-entries allows for the deployment of extended arrays of geophysical instruments that can fully capture the characteristics of objects analog to larger meteoroids. However, only five sample returns have happened since the end of the Apollo area: Genesis, Stardust, Hayabusa 1, Hayabusa 2, and OSIRIS-REx (Origins, Spectral Interpretation, Resource Identification, and Security-Regolith Explorer) \cite{Silber2023}.

NASA's OSIRIS-REx sample return mission visited 101955 Bennu, a small ($\sim$500$\,$m equatorial diameter) carbonaceous asteroid, to collect regolith samples. OSIRIS-REx was launched on 8 September 2016 and touched down on Bennu to successfully collect 121.6$\,$g of carbonaceous material on 20 October 2020. OSIRIS-REx left the asteroid on 10 May 2021 and returned its sample to Earth on 24 September 2023. The capsule, which had a maximum diameter of $\sim$80 cm and a height of $\sim$50 cm, reached the atmospheric interface at 14:42 UTC near the California coast before safely landing at 14:52 UTC at the Utah Test and Training Range in Dugway, Utah \cite[Figure \ref{fig1Map},][]{Francis2024}. 

Our study focuses on recording geophysical signals produced by the the flight of the OSIRIS-REx Sample Return Capsule (SRC) through the atmosphere using Distributed Acoustic Sensing (DAS), along with co-located seismometers and infrasound sensors.  

DAS is a laser-based technology that uses Rayleigh backscattering to detect distributed vibrations (e.g., strain rate) along the optical fibers. Previous studies have shown that a variety of seismoacoustic signals can be recorded with DAS, including seismic waves from earthquakes and explosions \cite[e.g.,][]{Fang2020, Lindsey2017}, T-phases from earthquakes \cite{Ugalde2021}, shock waves from thunder \cite{Zhu2019}, and meteorites \cite{VeraRodriguez2023}. Seismometers and infrasound sensors have been shown to record similar signals \cite[see reviews by][]{Campus2010,DannemannDugick2023}, and have been used to validate the accuracy and reliability of the strain and strain rate signals recorded by DAS in numerous settings \cite[e.g.,][]{Viens2024, Wang2018}. Previous sample return capsule re-entries have been recorded by seismometeters and infrasound sensors \cite[e.g.,][]{Edwards2007, ReVelle2007}; however, the OSIRIS-REx re-entry is the first for which DAS was deployed. 

\subsection*{Instrumentation and Data Collection}
\begin{figure}
    \centering
    \includegraphics[width=\textwidth,trim=0cm .2cm 0cm 0cm, clip]{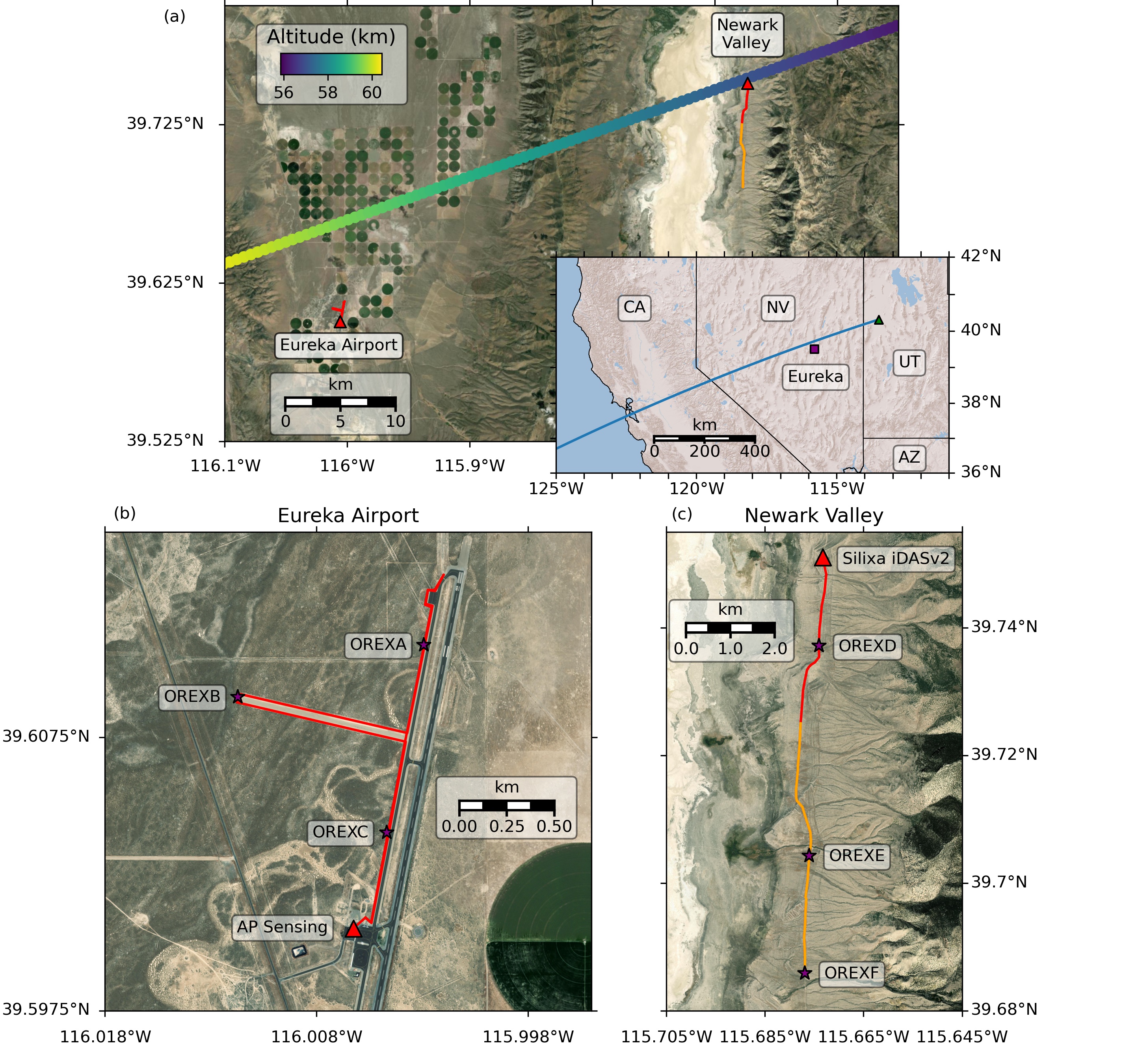}
    \caption{(a) Map of the region of interest including the two deployment sites: Eureka Airport and Newark Valley. The fiber-optic cable locations are shown by the colored lines with the color indicating the type of fiber (i.e., red is stainless steel jacketed cable and orange is tight buffered cable). The altitude of the OSIRIS-REx trajectory is also shown. The inset map shows the full re-entry trajectory of the sample return capsule over the western United States \cite[trajectory from][and see references within]{Francis2024, Silber2024}. (b) Map of the instruments at Eureka Airport. The location of the AP Sensing DAS interrogator is shown by the red triangle. The fiber runs from the DAS interrogator towards OREXC along the main taxiway, turns 90 degrees towards the west (towards OREXB) along the cross runway, then returns to the main taxiway and continues north towards OREXA. The fiber returns from the turnaround point to the north of OREXA and continues directly towards OREXC without returning to OREXB. (c) In Newark Valley, the fiber ran from the Silixa iDAS south past OREXD, OREXE, and OREXF.}
    \label{fig1Map}
\end{figure}

We deployed two sets of fiber-optic cables that were probed with two interrogators as well as six co-located seismometer/infrasound sensor pairs (OREXA, OREXB, OREXC, OREXD, OREXE, and OREXF), spread across two sites near Eureka, Nevada: at the Eureka Airport and in Newark Valley (Figure \ref{fig1Map}). Scientific and logistic feasibility influenced the choice of sites. Both sites were selected because they were located near the portion of the trajectory where the incoming capsule was anticipated to generate the strongest infrasound signals. The Newark Valley site was identified as ideal due to the availability of an extremely low-trafficked road stretching $>$7$\,$km in a heading sub-perpendicular to the trajectory that would allow any incoming signals produced by the re-entry to sweep along the fiber. The Eureka Airport, while further from the trajectory footprint, provided ease of access; the airport was also heavily instrumented by multiple institutions, and DAS data could be compared to other modalities. Instrument deployment is summarized below, for a more detailed description, including the multi-institution observation campaign to measure OSIRIS-REx SRC geophysical signals, see \cite{Silber2024}.  

A total of 12$\,$km of single mode fiber-optic cables were deployed across the two sites (Figure \ref{fig1Map}). An AP Sensing DAS interrogator (N5225B-R100) was used at the Eureka Airport site to probe 4.5$\,$km of stainless steel jacketed fiber and recorded data with a 500$\,$Hz sampling frequency, a 1.23$\,$m channel spacing, and a 5$\,$m gauge length. The cable was laid on the gravel that lined the paved runway. In Newark Valley, a Silixa iDASv2 interrogator (Version$\,$2.4.1.111) probed 7.5$\,$km of fiber: 3$\,$km of the stainless steel jacketed fiber (same fiber as at the airport site) spliced to 4.5$\,$km of tight buffered fiber-optic cable in a polyurethane jacket reinforced with aramid yarn (Figure S1). The iDASv2 interrogator recorded with a 500$\,$Hz sampling frequency, a 2$\,$m channel spacing, and a 10$\,$m gauge length. The cable was laid on a dirt road. Both DAS interrogators operated intermittently for testing purposes prior to the re-entry and then for at least 60 minutes prior to and 45 minutes after the expected re-entry. 

We installed seismometers (Geospace HS-1 3-component) and infrasound sensors (Hyperion 3000) at strategic locations near the optical fibers (see Table S1 and \cite{Silber2024} for further instrumentation details). The seismometers were buried sufficiently to cover the sensors, leveled, and oriented in the standard configuration (i.e., to true north). At each site, the infrasound sensor was placed on the ground as near as possible to the seismometer (10’s of cm), and a windscreen was placed over both sensors. All the instruments were placed within 5 m of the DAS fibers. The seismometers and infrasound sensors recorded with a 200$\,$Hz sampling rate and each seismometer-infrasound sensor pair was connected to a RefTek$\,$130 datalogger with timing information provided by Garmin GPS$\,$16x-HVS antennas. The seismometers and infrasound sensors were installed over several days prior to the re-entry and operated continuously until several hours after the re-entry.

\subsection*{Results}
\begin{figure}[htp]
    \centering
    \includegraphics[width=\textwidth,trim=0cm 0cm 0cm 0cm, clip]{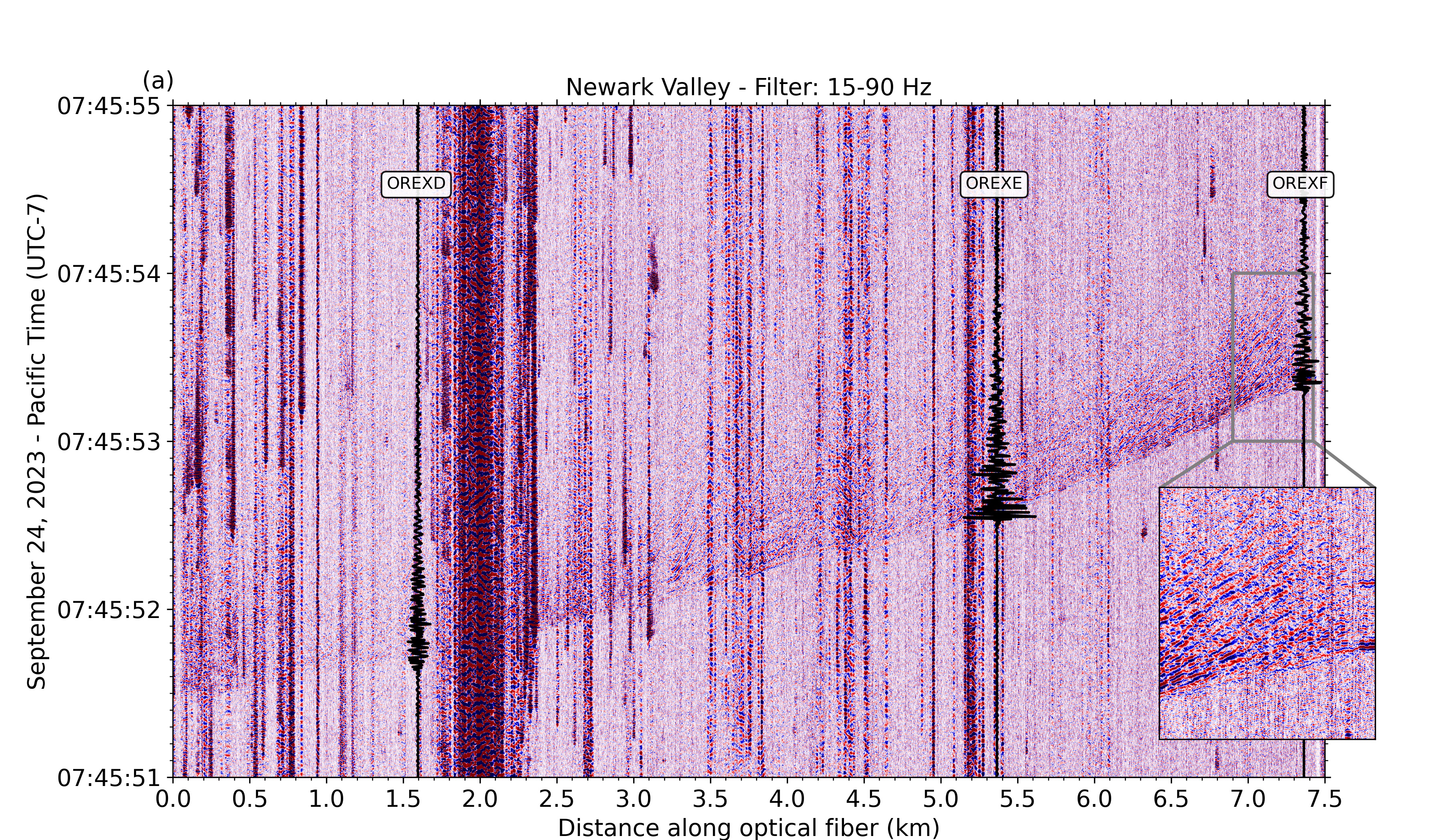}
    \includegraphics[width=\textwidth,trim=0cm 0cm 0cm 1cm, clip]{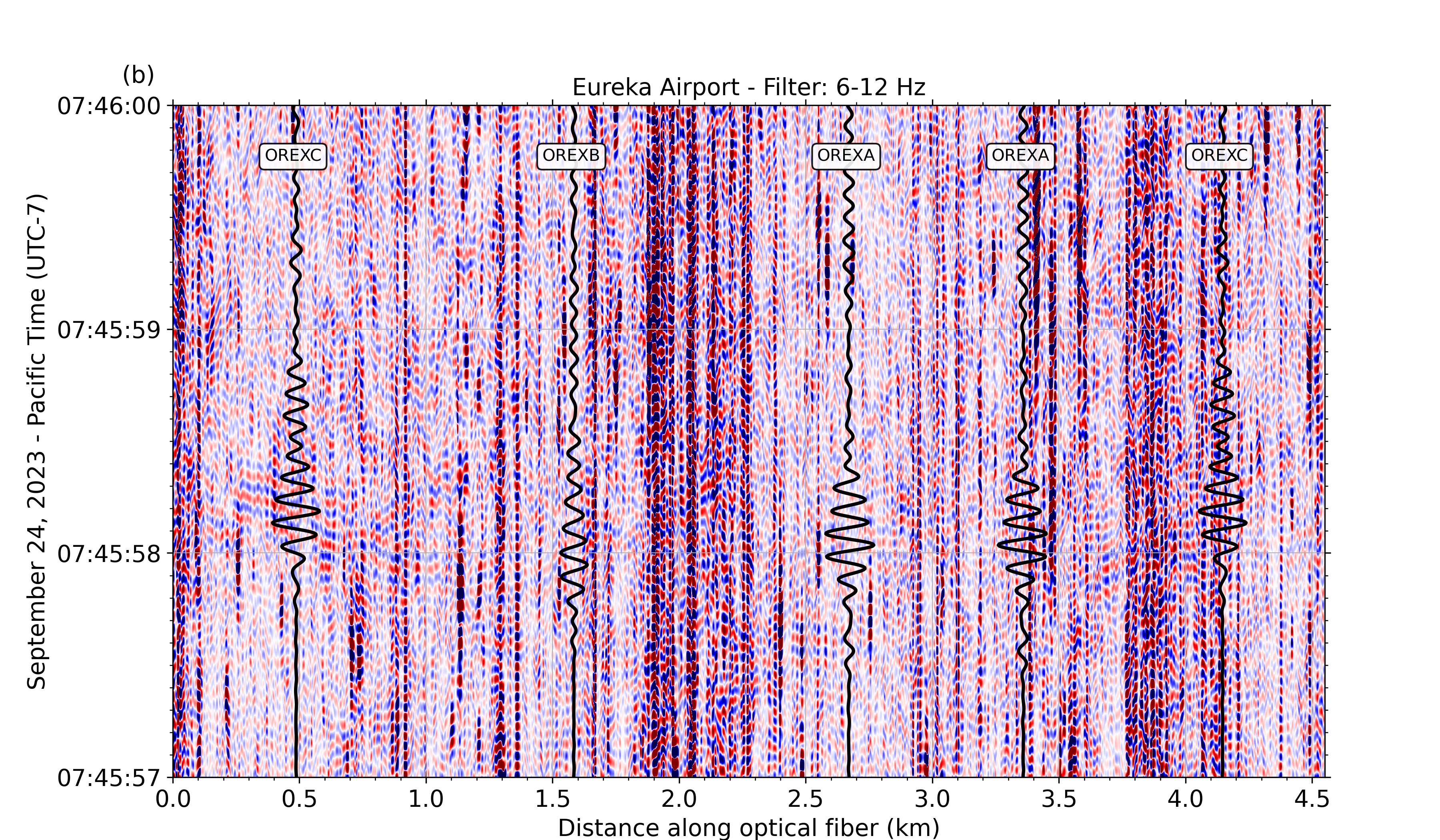}
    \caption{ Normalized strain-rate signals from the re-entry of the OSIRIS-REx sample return capsule recorded at (a) Newark Valley with a Silixa iDAS interrogator, and (b) the Eureka airport with an AP Sensing DAS interrogator. Normalized acceleration data from the north component at the six seismometers, which are obtained after taking the time derivative of the recorded velocity data, are overlaid on the plots. Different bandpass filters are applied to the Eureka airport and Newark Valley DAS datasets. The inset plot in (a) shows the zoom on secondary phases propagating along the cable after the initial arrivals.}
    \label{waterfall}
\end{figure}

\begin{figure}[htp]
    \centering
    \includegraphics[width=\textwidth,trim=4.5cm 1.2cm 4.5cm 2cm, clip]{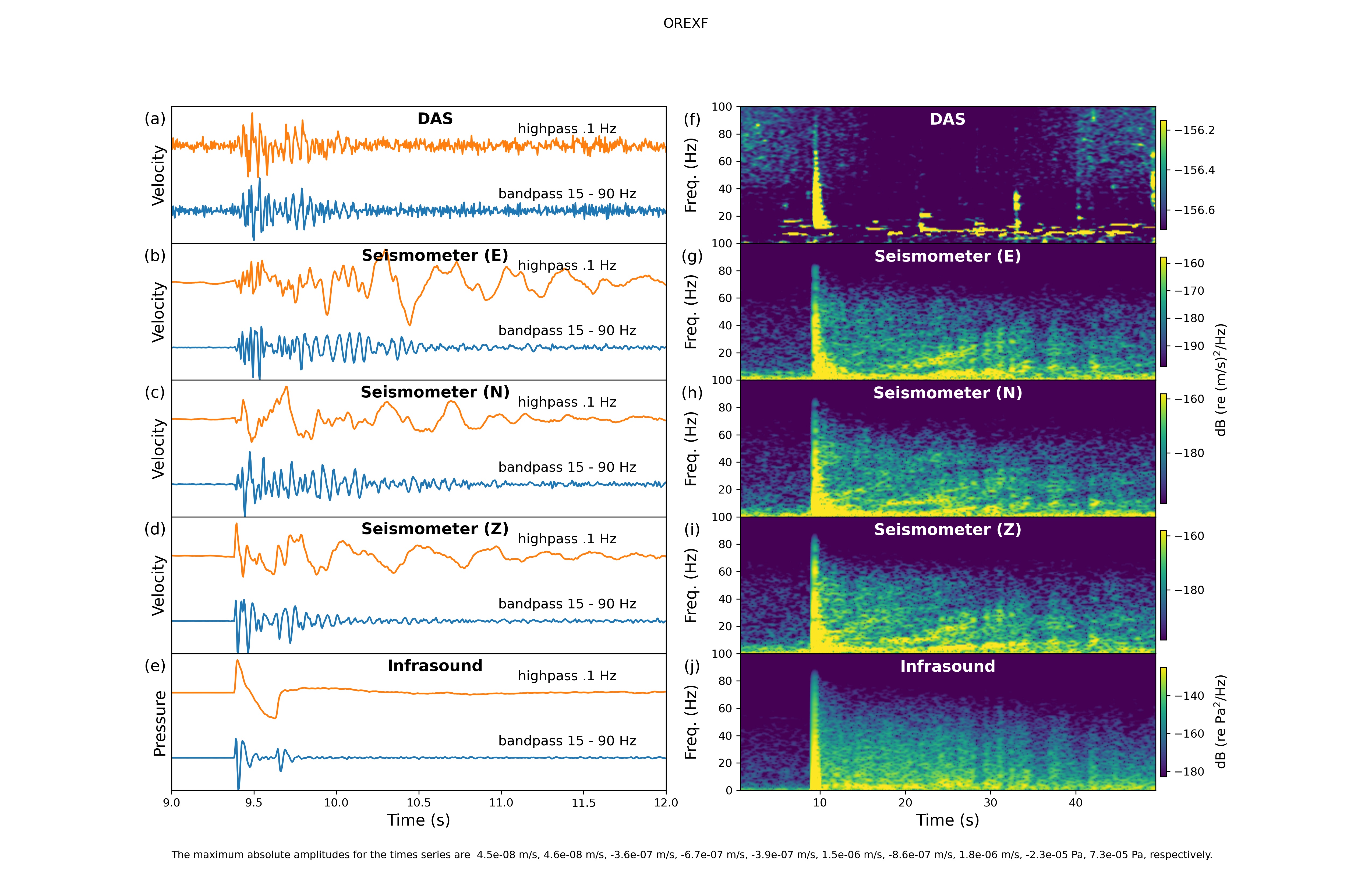}
    \caption{Time series (a--e) and spectrograms (f--j) of data for sensors at or near OREXF. DAS data were selected to be near OREXF and have suitable signal-to-noise (see Figure S2) and then stacked to further improve SNR. The maximum absolute amplitudes for the times series are: (a) 4.5e-08$\,$m/s (upper orange line) and 4.6e-08$\,$m/s (lower blue line) for the DAS; and (b) -3.6e-07$\,$m/s (upper orange line) and  -6.7e-07$\,$m/s (lower blue line),  (c) -3.9e-07$\,$m/s (upper orange line) and 1.5e-06$\,$m/s (lower blue line), and (d) -8.6e-07$\,$m/s (upper orange line) and 1.8e-06$\,$m/s (lower blue line) for the seismic data.}
    \label{waveformsSpecF}
\end{figure}

\begin{figure}[htp]
    \centering
    \includegraphics[width=\textwidth,trim=1cm 0cm 1cm 0cm, clip]{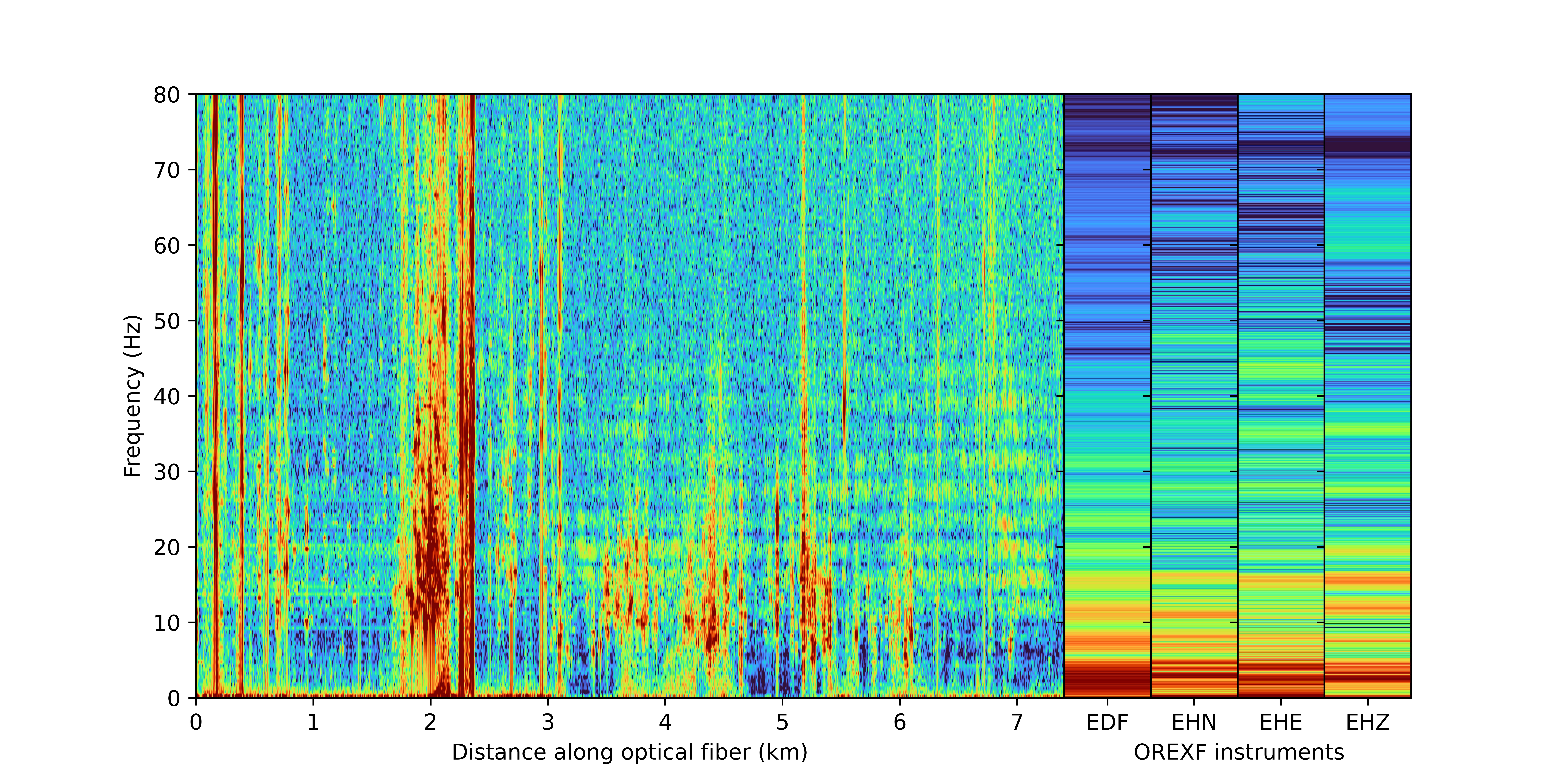}
    \caption{Spectra for all channels recorded with DAS at Newark Valley and corresponding OREXF seimo-acoustic instruments.}
    \label{spec}
\end{figure}

Figure \ref{waterfall} shows strain-rate data for both deployment locations over all DAS channels with overlays of the acceleration data from the co-located seismometers. In Figure \ref{waterfall}, the data collected at Newark Valley are bandpass filtered between 15 and 90$\,$Hz since there is little signal recorded on the DAS under 15 Hz (see Figures \ref{waveformsSpecF} and \ref{spec}). In Newark Valley (Figure \ref{waterfall}a), a clear signal can be observed along the entire fiber with the exception of a few noisy sections (e.g., around 2$\,$km from the interrogator). The onset time of the strain rate arrivals captured by the Silixa iDASv2 match well the seismometer-recorded arrival. The inset in Figure \ref{waterfall}a zooms in to show how the DAS data record well not only the first arrival, but also record different phases that would otherwise be difficult to distinguish with seismometers alone. The apparent velocity of these other phases is between 550$\,$m/s and 1000$\,$m/s.

At the Eureka Airport, a similar processing, including bandpass filtering between 15 and 90$\,$ Hz, did not reveal any identifiable arrival signals in the DAS data. However, when we apply a 6 to 12$\,$Hz bandpass filter and apply a moving median filter over 5 channels to increase the signal-to-noise ratio, weak arrivals are observed (Figure \ref{waterfall}b). The processed data show weak arrivals in the first 1,000$\,$m from the interrogator that match the arrival time of the seismometer data at OREXC. Note in both Figure \ref{waterfall}a and \ref{waterfall}b that the velocity data from all seismometers are converted to acceleration and bandpass filtered in the same frequency bands as the DAS data on the same subplots. In Figure \ref{waterfall}b, the OREXA and OREXC data are plotted twice with flipped polarities as the fiber path returned along the runway and passed by these two seismometers twice (see Figure \ref{fig1Map}c). 

The discrepancy of the data quality at the two locations is likely due to the distance from the source and the relative amounts of anthropogenic noise at the airport site near a major road. Newark Valley was closer to the source signal generation as evidenced by the earlier arrival time. At Newark Valley, the signal-to-noise of the infrasound and the seismometers were approximately 3.8 and 1.8 times higher, respectively, than at the Eureka airport. For the remainder of the analysis, we therefore focus on the Newark Valley data, particularly near seismometer/infrasound OREXF where the arrival was prominent in the DAS data. 

It is a common practice to use co-located seismometers to validate the amplitude and phase information of DAS data \cite[e.g.,][]{Lindsey2020, Wang2018}. However, since the fiber was surface draped, it was in contact with both the ground and the air. Therefore, we aim to discern the degree to which the DAS measurements are composed of signal that is from the direct wave traveling through the air to the fiber versus the wave impacting the ground and converting to a seismic signal (e.g., ground-coupled airwaves). First, to compare the DAS data recorded in Newark Valley with the seismometer and infrasound data, we converted the strain rate data produced by the Silixa iDASv2 to velocity following the methodology proposed by \cite{Trabattoni2023}. Due to poor coupling, sections of the fiber under tension, wind noise, optical fiber construction, and orientation to the sample return capsule trajectory, a large variation is observed in the unwanted noise that obscures the signal generated by the sample return capsule. Consequently, only channels with high signal-to-noise ratio (SNR) near the co-located seismoacoustic sensors are selected for further comparative analysis.  We define high-SNR channels as those where the standard deviation of the bandpass filtered signal (15 Hz - 55$\,$Hz) during an expected arrival window (width 1$\,$s) exceeds a threshold of 1.7 relative to a similarly processed 1$\,$s window preceding the expected arrival. The approximate arrival times on individual DAS channels were determined by interpolating based on a parabolic fit on the waterfall plot in Figure \ref{waterfall}a, using the infrasound arrivals as control points. DAS channels were considered near a seismoacoustic station if they were within 720 m. A plot showing the selected high-SNR DAS sensors can be found in Figure S2. 

Fig. \ref{waveformsSpecF} shows comparative time series and spectrograms for all sensing modalities near OREXF (high-SNR DAS channels within 720$\,$m of OREXF, OREXF seismometer, and OREXF infrasound sensor). The complete set of figures corresponding to all Newark Valley seismoacoustic stations can be found in the supplementary materials (Figures S3 and S4), and we plot the waveforms from all six seismic and infraound stations in Figures S5 and S6. In Figure \ref{waveformsSpecF}a, the time series of the DAS data were downsampled to 200 Hz and stacked to enhance the signal. In Figure \ref{waveformsSpecF}f, spectrograms of individual DAS sensors were calculated first prior to stacking. It is clear from the spectrograms that the DAS was not sensitive to the low frequency signal generated by the sample capsule return, consequently, the lower plots in Figure \ref{waveformsSpecF}a--e also show signals that have been bandpassed between 15--90 Hz. The DAS time series shows features of both the seismometer and infrasound time series. The 15--90$\,$Hz bandpassed DAS timeseries partially captures the extended coda that is pronounced in the bandpassed seismometer data. Including the lower frequency content (highpass 0.1$\,$Hz time series), we observe a clear N-wave arrival in the infrasound time series (Figure \ref{waveformsSpecF}e), and the timing of the N-wave maximum and minimum amplitudes correspond to the two packets of higher amplitude signal in the highpass DAS timeseries in Figure \ref{waveformsSpecF}a. 

To explore the frequency content further, we plot the normalized spectrum for all Newark Valley DAS channels in Figure \ref{spec}. For comparison, the right side of the plot also shows the spectra for the seismoacoustic data at OREXF. A considerable amount of noise obscures the DAS signal, however, there are clearly horizontal stripes that extend almost the entire length of the optical fiber. The precise frequency of these stripes corresponds particularly well to the infrasound data (EDF), but not as well to the seismic data. An N-wave exhibits evenly spaced peaks in the frequency domain, therefore, it is likely the DAS is capturing the characteristic signals from the overpressure profile associated with the sonic boom.

\begin{figure}[htp]
    \centering
    \includegraphics[width=\textwidth,trim=0cm 0cm 0cm 0cm, clip]{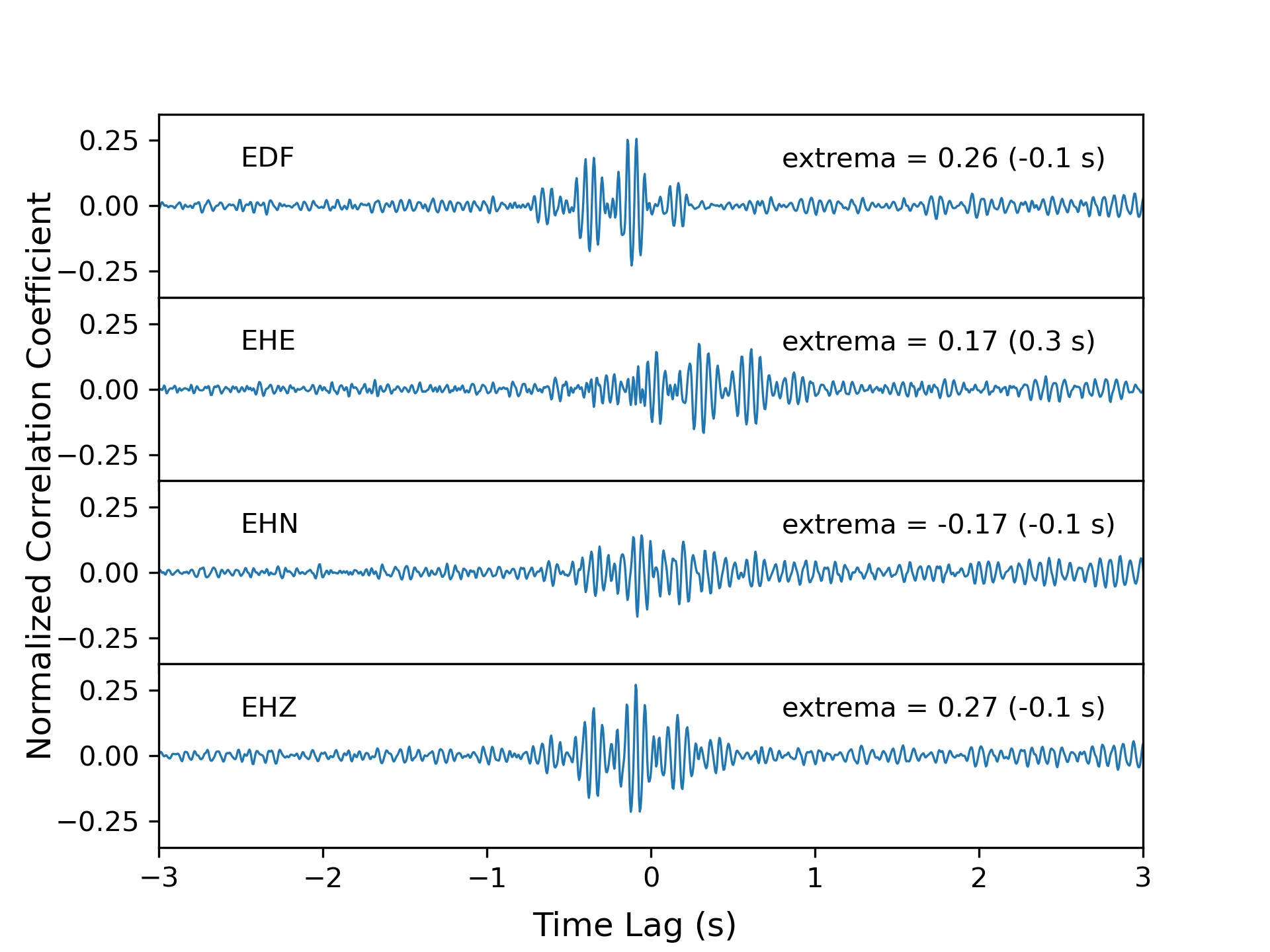}
    \caption{Cross correlation plots, each of the seismoacoustic channels from location OREXF is cross-correlated with the high-SNR, stacked DAS data from near OREXF (i.e., the lower line in Figure \ref{waveformsSpecF}a), using a 15--90$\,$Hz bandpass filter.}
    \label{crossCorr}
\end{figure}

Finally, normalized cross correlations between the stacked DAS data near OREXF and the OREXF seismoacoustic data (bandpass filtered between 15 and 90 Hz) are shown in Figure \ref{crossCorr}. Surprisingly, the highest coefficient was calculated for the vertical component of the seismic sensor (EHZ), with the infrasound being slightly smaller. Given that the optical fiber in Newark Valley is primarily oriented north-south, we would have expected the north (EHN) component to have a larger correlation coefficient. The horizontal components have the smallest correlation coefficients. For the infrasound, and the north and vertical seismic channels, we observe that peak correlation is 0.1$\,$s early compared with the DAS arrivals, while the east seismic channel peak correlation lags behind the DAS arrivals by 0.3$\,$s. These lags are notable given the signal duration, and we suspect that the DAS cable physical response may influence the timing delays we observe. 

\subsection*{Discussion}
Together, these results demonstrate the potential of DAS for measuring signals from hypersonic/supersonic sources traveling through the atmosphere, including bolides and their analogs. In particular, the high spatial density of DAS sensors allows us to record the full wavefield and visualize phases that would be otherwise missed or more difficult to interpret with a traditional, sparse seismoacoustic deployment. 

At Newark Valley, two different optical fiber cables that have both been designed for use with DAS interrogators were utilized, one cable with a steel jacket and another tight buffered. We found that the quality of the signal depended strongly on the cable construction, with the first 3 km of steel jacketed cable being fairly insensitive to the signal from the space capsule while still recording high levels of wind noise. It is noted that the signal quality does improve further away from the source, which is attributed to the angle of incidence relative to the cable. However, especially in Figure \ref{spec}, there is a distinct transition at the 3 km mark where the two cables meet. When we were placing the cable on the ground surface, we observed that we could hear the steel cable transmitting sound from far away, such as a person walking near the cable, and that the material construction could be causing the cable itself to serve as an acoustic waveguide. However, both cables were designed for burial, so it is possible that our observations would not be consistent with other deployment scenarios. Noise from wind was difficult to mitigate as the broadband frequency of a gust of wind overlapped strongly with the signals from the space capsule. In the days leading up to the re-entry, we discovered that even placing the cables close to, but not in, the short vegetation near the dirt road produced more pronounced wind noise. Initially, we had placed the cable at the edge of the road near the vegetation since the road had occasional traffic, but based on the observed noise, we moved the fiber away from the vegetation and directly onto the dirt road the evening before the re-entry (Figure S7). 

The absence of low-frequency content at Newark Valley is perplexing, as the Silixa iDAS is routinely used to measure signals in that bandwidth. It is possible that deploying it on the surface rather than burying could have contributed to the observed lack of low-frequency content, but the physical mechanism is unknown. A handful of useful papers have explored surface draped fiber \cite{Harmon2022, Pandey2023, Spikes2019}, but generally they have employed active seismic sources with central frequencies higher than 15 Hz and large differences in the frequency content were observed for different cables, therefore this lack of low frequency content has not been particularly noted in other surface-draped experiments. 

\subsection*{Conclusions}
We successfully measured seismoacoustic signals produced by the OSIRIS-REx Sample Return Capsule using DAS, representing the first such measurement with this instrumentation. The recordings highlight the strength of DAS in terms of sampling the continuous evolution of the wavefield over scales of 100's to several 1,000's of meters. We observed initial impulsive arrivals, with strong coda lasting about a second. Because of the inherent spatial sampling density of DAS, we were able to observe phases that would have been missed otherwise by a standard, more spatially sparse seismoacoustic deployment. The co-located seismometers and infrasound sensors provided context to help interpret the variety of phases recorded by the DAS, indicating that the fiber recorded a combination of both broadband acoustic and seismic phases.  We observe impacts of deployment and environmental conditions on data quality --- in particular we observed strong contrasts in SNR between different cable types.

Our research can be used to inform future deployment choices. The surface-draped installation clearly resulted in higher noise conditions than a trenched installation would have, but we were still able to record the re-entry signal. Trenching will not always be feasible for all DAS deployments, whether for environmental or logistical reasons, and it is therefore important to understand how to optimize recordings in these conditions. One of our ongoing objectives is to better understand the physical mechanisms impacting the ground-air coupling of acoustic and seismic phases as they are recorded with surface-draped DAS fiber. 
 
\subsection*{Data and Resources}
Data are available from the authors upon request. The supplemental material consists of one table and seven figures.

\subsection*{Declaration of Competing Interests}
The authors acknowledge that there are no conflicts of interest recorded.

\subsection*{Acknowledgements}
This research was supported by Los Alamos National Laboratory (LANL) through the Laboratory Directed Research and Development (LDRD) program, under project number 20220188DR and under the LANL Center for Space and Earth Science (CSES) LDRD project number 20240477CR-SES with publication number LA-UR-24-30604. L.V. was partially supported by the CSES Chick Keller Fellowship. Additional funding for E.A.M. was provided by the Pat Tillman Foundation Scholarship and the Colorado State University Vice President for Research Graduate Fellowship Program. We thank the Eureka County Commission and Eureka Airport Manager Jayme Halpin for access at the airport site, and the Bristlecone Bureau of Land Management Field Office for guidance in complying with casual use requirements at the Newark Valley site.  We are grateful for the loan of field vehicles from the MPA-Q Division and EES-14 Group at LANL.  We thank Elizabeth Silber, Philip Blom, Jordan Bishop, Christopher Johnson, and Jeremy Webster for the helpful discussions that informed our analysis, and to Jeremy Webster for additional assistance with instrument preparation. We thank Elizabeth Silber, Daniel Bowman, and Michael Fleigle for inviting C.G.C. on a scouting trip to confirm site feasibility.

\printbibliography

@string{atm = {Atmosphere}}

@string{bssa = {Bull. Seismol. Soc. Am.}}

@string{csr = {Chem. Soc. Rev.}}

@string{geophs = {Geophysics}}

@string{gji = {Geophys. J. Int.}}

@string{grl = {Geophys. Res. Lett.}}

@string{jgr = {J. Geophys. Res.}}

@string{mps = {Meteorit. Planet. Sci.}}

@string{nat = {Nature}}

@string{nsfcgeo = {Near Surf. Geophys.}}

@string{psj = {Planet. Sci. J.}}

@string{sci = {Science}}

@string{srl = {Seismol. Res. Lett.}}

@string{ssr = {Space Sci. Rev.}}

@article{Brown2013,
    author = {Brown, P. G. and Assink, J. D. and Astiz, L. and Blaauw, R. and Boslough, M. B. and Borovi\u{c}ka, J. and Brachet, N. and Brown, D. and Campbell-Brown, M. and Ceranna, L. and Cooke, W. and de Groot-Hedlin, C. and Drob11, D. P. and Edwards, W. and Evers, L. G. and Garces, M. and Gill, J. and Hedlin, M. and Kingery, A. and Laske, G. and Le Pichon, A. and Mialle, P. and Moser, D. E. and Saffer, A. and Silber, E. and Smets, P. and Spalding, R. E. and Spurny, P. and Tagliaferri, E. and Uren, D. and Weryk, R. J. and Whitaker, R. and Krzeminski, Z.},
    title = {{A 500-kiloton airburst over Chelyabinsk and an enhanced hazard from small impactors}}, 
    journal = nat,
    volume = {503},
    pages = {238--241},
    doi = {10.1038/nature12741}, 
    year = {2013}
}

@inbook{Campus2010,
    author = {Campus, P. and Christie, D.R},
    title = {Worldwide Observations of Infrasonic Waves},  
    booktitle = {Infrasound Monitoring for Atmospheric Studies},
    publisher = {Springer Dordrecht},
    pages = {185--234},
    doi = {10.1007/978-1-4020-9508-5_6},
    year = {2010}
}

@article{Ceplecha1998,
    author = {Ceplecha, Zden\u{e}k and Borovi\u{c}ka, Ji\u{r}\'{I} and Elford, W. Graham and ReVelle, Douglas O. and Hawkes, Robert L. and Porub\u{c}an, Vladim\'{I}r and \u{S}imek, Milo\u{s}},
    title = {{Meteor phenomena and bodies}},
    journal = ssr,
    volume = {84},
    pages = {327--471},
    doi = {10.1023/A:1005069928850},
    year = {1998}
}

@article{DannemannDugick2023,
    author = {Dannemann Dugick, Fransiska and Koch, Clinton and Berg, Elizabeth and Arrowsmith, Stephen and Albert, Sarah},
    title ={{A new decade in seismoacoustics (2010–2022)}},
    journal = bssa, 
    volume = {113}, 
    number = {4},
    pages = {1390--1423}, 
    doi = {10.1785/0120220157},
    year = {2023}
}

@article{Edwards2007,
	author = {Edwards, Wayne N. and Eaton, David W. and McCausland, Philip J. and ReVelle, Douglas O. and Brown, Peter G.},
    title = {{Calibrating infrasonic to seismic coupling using the Stardust sample return capsule shockwave: Implications for seismic observations of meteors}},	
    journal = jgr,
	volume = {112},
    number = {10},
    pages = {13 pp.},
    doi = {10.1029/2006JB004621},
	year = {2007}
}

@article{Fang2020,
    author = {Fang, Gang and Li, Yunyue Elita and Zhao, Yumin and Martin, Eileen R.},
    title = {{Urban near-surface seismic monitoring using distributed acoustic sensing}},
    journal = grl,
    volume = {47}, 
    number = {6}, 
    pages = {9 pp.},
    doi = {10.1029/2019GL086115},
    year = {2020}
}

@article{Francis2024,
	author = {Francis, S. R. and Johnson, M. A. and Queen, E. and Williams, R. A.},
    title = {Entry, Descent, and Landing Analysis for the OSIRIS-REx Sample Return Capsule},
    journal = {46th Annual AAS Guidance, Navigation and Control (GN\&C) Conference Program Papers},
    volume = {24}, 
    number = {188},
	pages = {22 pp.},
    doi = {},
	year = {2024}
}

@article{Harmon2022,
    author={Harmon, Nicholas and Rychert, Catherine A and Davis, John and Brambilla, Gilberto and Buffet, William and Chichester, Ben and Dai, Yuhang and Bogiatzis, Petros and Snook, James and van Putten, Lieke and others},  
    title={{Surface deployment of DAS systems: Coupling strategies and comparisons to geophone data}},
    journal= nsfcgeo,
    volume={20},
    number={5},
    pages={465--477},
    doi = {10.1002/nsg.12232},
    year={2022},
}

@article{Lindsey2017, 
    author = {Lindsey, Nathaniel J. and Martin, Eileen R. and Dreger, Douglas S. and Freifeld, Barry and Cole, Stephen and James, Stephanie R. and Biondi, Biondo L. and Ajo-Franklin, Jonathan B.},
    title = {{Fiber-optic network observations of earthquake wavefields}}, 
    journal = grl,
    volume = {44}, 
    number = {23}, 
    pages = {11792--11799}, 
    doi = {10.1002/2017GL075722},
    year = {2017}
}

@article{Lindsey2020,
    author = {Lindsey, Nathaniel J. and Rademacher, Horst and Ajo-Franklin, Jonathan B.},
    title = {On the Broadband Instrument Response of Fiber-Optic DAS Arrays},
    journal = jgr,
    volume = {125},
    number = {2},
    pages = {16 pp.},
    doi = {10.1029/2019JB018145},
    year = {2020}
}

@article{Pandey2023,
    author={Pandey, Adesh and Shragge, Jeffrey and Chambers, Derrick and Girard, Aaron J},  
    title={{Developing a distributed acoustic sensing seismic land streamer: Concept and validation}},
    journal=geophs,
    volume={88},
    number={6},
    pages={WC59--WC67},
    doi = {10.1190/geo2023-0072.1},
    year={2023},
}

@article{Plane2012,
    author = {Plane, John M. C.},
    title = {{Cosmic dust in the earth's atmosphere}},
    journal = csr,
    volume = {41},
    pages = {6507--6518},
    doi = {10.1039/C2CS35132C},
    year = {2012}
}

@article{Popova2013,
    author = {Popova, Olga P. and Jenniskens, Peter and Emel’yanenko, Vachelav and Kartashova, Anna and Biryukov, Eugeny and Khaibrakhmanov, Sergey and Shuvalov, Valery and Rybnov, Yurij and Dudorov, Alexandr and Grokhovsky, Victor I. and Badyukov, Dmitry D. and Yin, Qing-Zhu and Gural, Peter S. and Albers, Jim and Granvik, Mikael and Evers, L{\"{a}}slo G. and Kuiper, Jacob and Kharlamov, Vladimir and Solovyov, Andrey and Rusakov, Yuri S. and Korotkiy, Stanislav and Serdyuk, Ilya and Korochantsev, Alexander V. and Larionov, Michail Yu. and Glazachev, Dmitry and Mayer, Alexander E. and Gisler, Galen and Gladkovsky, Sergei V. and Wimpenny, Josh and Sanborn, Matthew E. and Yamakawa, Akane and Verosub, Kenneth L. and Rowland, Douglas J. and Roeske, Sarah and Botto, Nicholas W. and Friedrich, Jon M. and Zolensky, Michael E. and Le, Loan and Ross, Daniel and Ziegler, Karen and Nakamura, Tomoki and Ahn, Insu and Lee, Jong Ik and Zhou, Qin and Li, Xian-Hua and Li, Qiu-Li and Liu, Yu and Tang, Guo-Qiang and Hiroi, Takahiro and Sears, Derek and Weinstein, Ilya A. and Vokhmintsev, Alexander S. and Ishchenko, Alexei V. and Schmitt-Kopplin, Phillipe and Hertkorn, Norbert and Nagao, Keisuke and Haba, Makiko K. and Komatsu,Mutsumi and Mikouchi, Takashi and {the Chelyabinsk Airburst Consortium}},
    title = {{Chelyabinsk airburst, damage assessment, meteorite recovery, and characterization}},
    journal = sci, 
    volume = {342},
    number = {6162},
    pages = {1069--1073},
    doi = {10.1126/science.1242642},
    year = {2013}
}

@article{ReVelle2007,
	author = {ReVelle, D. O. and Edwards, W. N.},
    title = {{Stardust --- An artificial, low-velocity "meteor" fall and recovery: 15 January 2006}},
	journal = mps,
    volume = {42},
	number = {3},
	pages = {271--299},
    doi = {10.1111/j.1945-5100.2007.tb00232.x},
	year = {2007}
}

@article{Silber2023, 
    author = {Silber, Elizabeth A. and Bowman, Daniel C. and Albert, S.}, 
    title = {{A review of infrasound and seismic observations of sample return capsules since the end of the apollo era in anticipation of the OSIRIS-REx arrival}}, 
    journal = atm, 
    volume = {14}, 
    number = {10},
    pages = {19 pp.}, 
    doi = {10.3390/atmos14101473}, 
    year = {2023}
}

@article{Silber2024, 
    author = {Silber, Elizabeth A. and Bowman, Daniel C. and Carr, Chris G. and Eisenberg, David P. and Elbing, Brian R. and Fernando, Benjamin and Garc\'{e}s, Milton A. and Haaser, Robert and Krishnamoorthy, Siddharth and Langston, Charles A. and Nishikawa, Yasuhiroa and Webster, Jeremy and Anderson, Jacob F. and Arrowsmith, Stephen and Bazargan, Sonia and Beardslee, Luke and Beck, Brant and Bishop, Jordan W. and Blom, Philip and Bracht, Grant and Chichester, David L. and Christe, Anthony and Cummins, Kenneth and Cutts, James and Danielson, Lisa and Donahue, Carly and Eack, Kenneth and Fleigle, Michael and Fox, Douglas and Goel, Ashish and Green, David and Hasumi, Yuta and Hayward, Chris and Hicks, Dan and Hix, Jay and Horton, Stephen and Hough,Emalee and Huber, David P. and Hunt, Madeline A. and Inman,Jennifer and Islam, S. M. Ariful and Izraelevitz, Jacob and Jacob, Jamey D. and Clarke, Jacob and Johnson, James and KC, Real J. and Komjathy, Attila and Lam, Eric and LaPierre, Justin and Lewis, Kevin and Lewis, Richard D. and Liu, Patrick and Martire, L\'{e}o and McCleary, Meaghan and McGhee, Elisa A. and Mitra, Ipsita and Nag, Amitabh and Ocampo Giraldo, Luis and Pearson, Karen and Plaisir, Mathieu and Popenhagen, Sarah K. and Rassoul, Hamid and Ronac Giannone, Miro and Samnani, Mirza and Schmerr, Nicholas and Spillman, Kate and Srinivas, Girish and Takazawa, Samuel K. and Tempert, Alex and Turley, Reagan and Van Beek, Cory and Viens, Lo\"{i}c and Walsh, Owen A. and Weinstein, Nathan and White, Robert and Williams, Brian and Wilson, Trevor C. and Wyckoff, Shirin and Yamamoto, Masa-yuki and Yap, Zachary and Yoshiyama, Tyler and Zeiler, Cleat},
    title = {{Geophysical observations of the 24 September 2023 OSIRIS-REx sample return capsule re-entry}},
    journal = psj, 
    volume = {5}, 
    number = {213},
    pages = {47 pp.},
    doi = {10.3847/PSJ/ad5b5e},
    year= {2024}
}

@article{Spikes2019, 
    author={Spikes, Kyle T and Tisato, Nicola and Hess, Thomas E and Holt, John W},
    title={{Comparison of geophone and surface-deployed distributed acoustic sensing seismic data}},
    journal=geophs,
    volume={84},
    number={2},
    pages={A25--A29},
    doi = {10.1190/geo2018-0528.1},
    year={2019},
}

@article{Trabattoni2023,
    author={Trabattoni, Alister and Biagioli, Francesco and Strumia, Claudio and van den Ende, Martijn and Scotto di Uccio, Francesco and Festa, Gaetano and Rivet, Diane and Sladen, Anthony and Ampuero, Jean Paul and M{\'e}taxian, Jean-Philippe and others},  
    title={{From strain to displacement: Using deformation to enhance distributed acoustic sensing applications}},
    journal= gji, 
    volume={235},
    number={3},
    pages={2372--2384},
    doi = {10.1093/gji/ggad365},
    year={2023},
}

@article{Ugalde2021,
	author = {Ugalde, Arantza and Becerril, Carlos and Villase{\~n}or, Antonio and Ranero, C{\'e}sar R. and Fern{\'a}ndez‐Ruiz, Mar{\'\i}a R. and Martin‐Lopez, Sonia and Gonz{\'a}lez‐Herr{\'a}ez, Miguel and Martins, Hugo F.},
    title = {{Noise levels and signals observed on submarine fibers in the Canary Islands using DAS}},	
    journal = srl,
    volume = {93},
	number = {1},
	pages = {351--363},
    doi = {10.1785/0220210049},
	year = {2021}
}

@article{VeraRodriguez2023,
	author = {{Vera Rodriguez}, Ismael and Isken, Marius P. and Dahm, Torsten and Lamb, Oliver D. and Wu, Sin-Mei and Kristj{\'{a}}nsd{\'{o}}ttir, Sigr{\'{i}}our and J{\'{o}}nsd{\'{o}}ttir, Krist{\'{i}}n and Sanchez-Pastor, Pilar and Clinton, John and Wollin, Christopher and Baird, Alan F. and Wuestefeld, Andreas and Booz, Beat and Eibl, Eva P. S. and Heimann, Sebastian and Goertz-Allmann, Bettina P. and Jousset, Philippe and Oye, Volker and Hj{\"{o}}rleifsd{\'{o}}ttir, Vala and Obermann, Anne},
    title = {{Acoustic signals of a meteoroid recorded on a Large-N seismic network and fiber-optic cables}},
	journal = srl,
    volume = {94},	
    number = {2 A},
	pages = {731--745},
    doi = {10.1785/0220220236},
	year = {2023}
}

@article{Viens2024,
	author = {Viens, Lo\"{i}c and Delbridge, Brent},
	title = {{Shallow soil response to a buried chemical explosion with geophones and distributed acoustic sensing}},
    journal = jgr,
    volume = {129}, 
    number = {7},
    pages = {22 pp.},
    doi = {10.1029/2023JB028416},
	year = {2024}
}

@article{Wang2018,
	author = {Wang, Herbert F and Zeng, Xiangfang and Miller, Douglas E and Fratta, Dante and Feigl, Kurt L and Thurber, Clifford H and Mellors, Robert J},
    title = {{Ground motion response to an ML 4.3 earthquake using co-located distributed acoustic sensing and seismometer arrays}},	
    journal = gji,
    volume = {213},
	number = {3},
	pages = {2020--2036},
	doi = {10.1093/gji/ggy182},
	year = {2018}
}

@article{Zhu2019, 
    author = {Zhu, Tieyuan and Stensrud, David J.},
    title = {{Characterizing thunder‐induced ground motions using fiber optic distributed acoustic sensing array}}, 
    journal = jgr,
    volume = {124},
    number = {23},
    pages = {12,810--12,823}, 
    doi = {10.1029/2019JD031453}, 
    year = {2019}
}

\clearpage

\end{document}


\begin{center}
    {\textbf{Supplemental material for:\\ Detection of a space capsule entering earth's atmosphere with distributed acoustic sensing (DAS)}}

Chris G. Carr, Carly M. Donahue\footnote{Corresponding author: cmd@lanl.gov}, Lo\"{i}c Viens, Luke B. Beardslee, Elisa A. McGhee, Lisa R. Danielson
\end{center}

\begin{figure*}[htp]
    \centering
    \includegraphics[width=\textwidth,trim=0.0cm 0cm 0.0cm 0.0cm, clip]{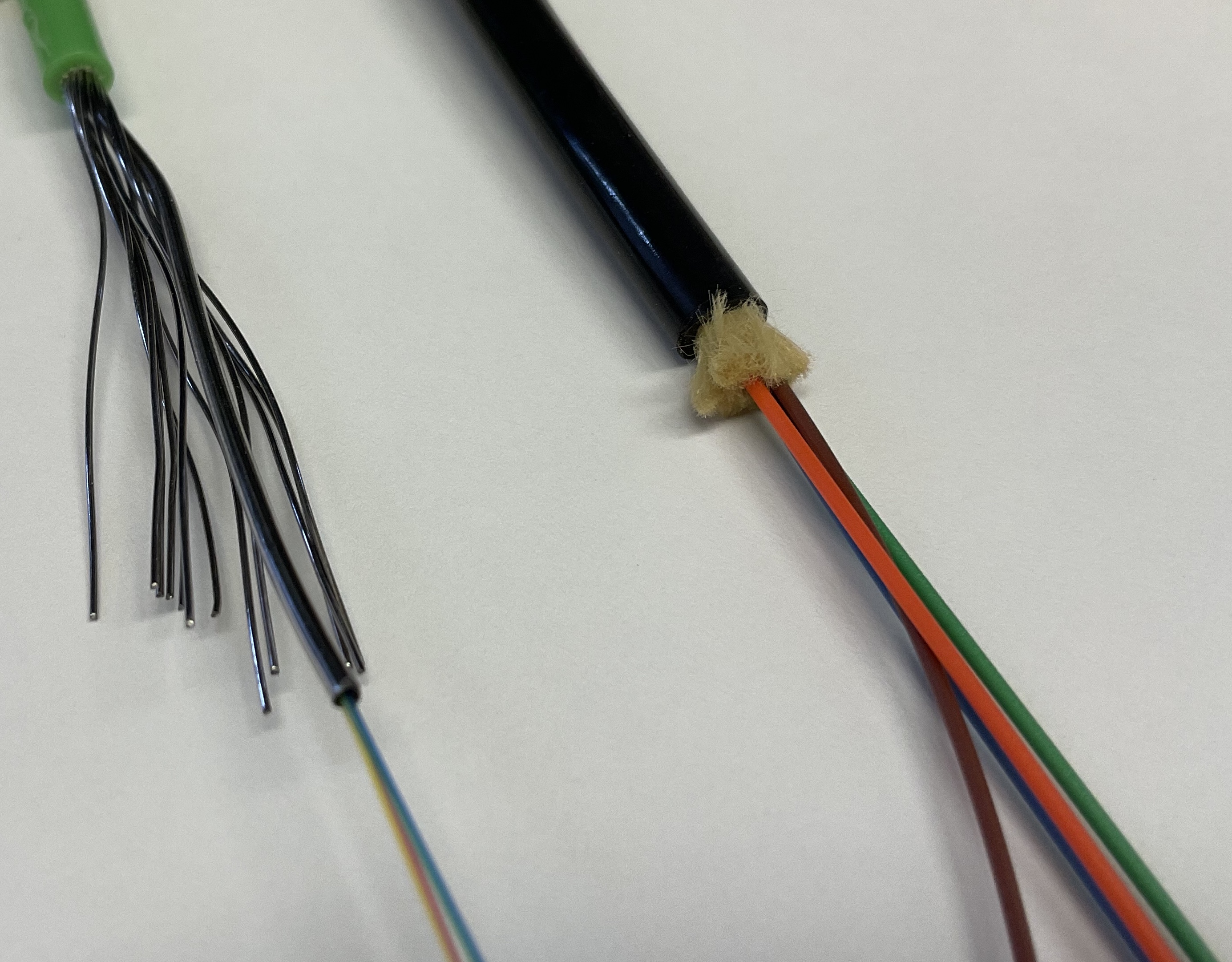}
    \caption{The two different fiber-optic cables that were used. The fiber on the left shows the stainless steel jacketed fiber and one on the right shows the tight buffered fiber-optic cable in a polyurethane jacket reinforced with aramid yarn.}
    \label{SuppDASCablePhoto}
\end{figure*}

\begin{figure*}[htp]
    \centering
    \includegraphics[width=\textwidth,trim=1.2cm 0cm 1.2cm 1cm, clip]{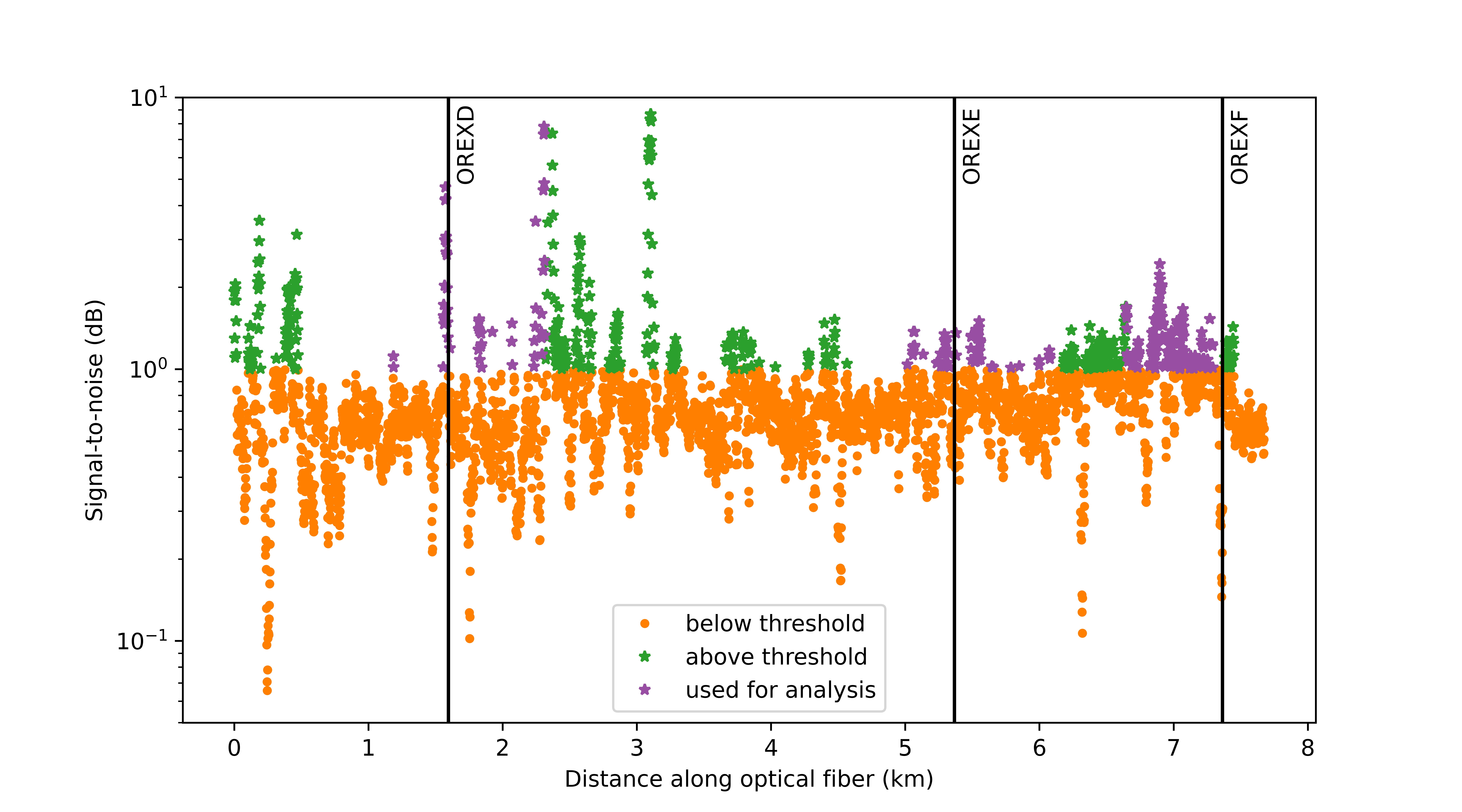}
    \caption{Using the standard deviation of the bandpass-filtered signal (15 Hz - 55$\,$Hz) for the DAS data, we calculate the SNR of a 1$\,$ second window during the expected arrival time relative to that of the preceding second. We used a threshold of 1.7, with channels below this threshold plotted in orange. For channels with SNR above the threshold, we selected channels within 720$\,$m of a seismoacoustic station for further comparison (purple).}
    \label{SuppNoiseCriterion}
\end{figure*}

\begin{figure*}[htp]
    \centering
    \includegraphics[width=\textwidth,trim=4.5cm 1.2cm 4.5cm 2cm, clip]{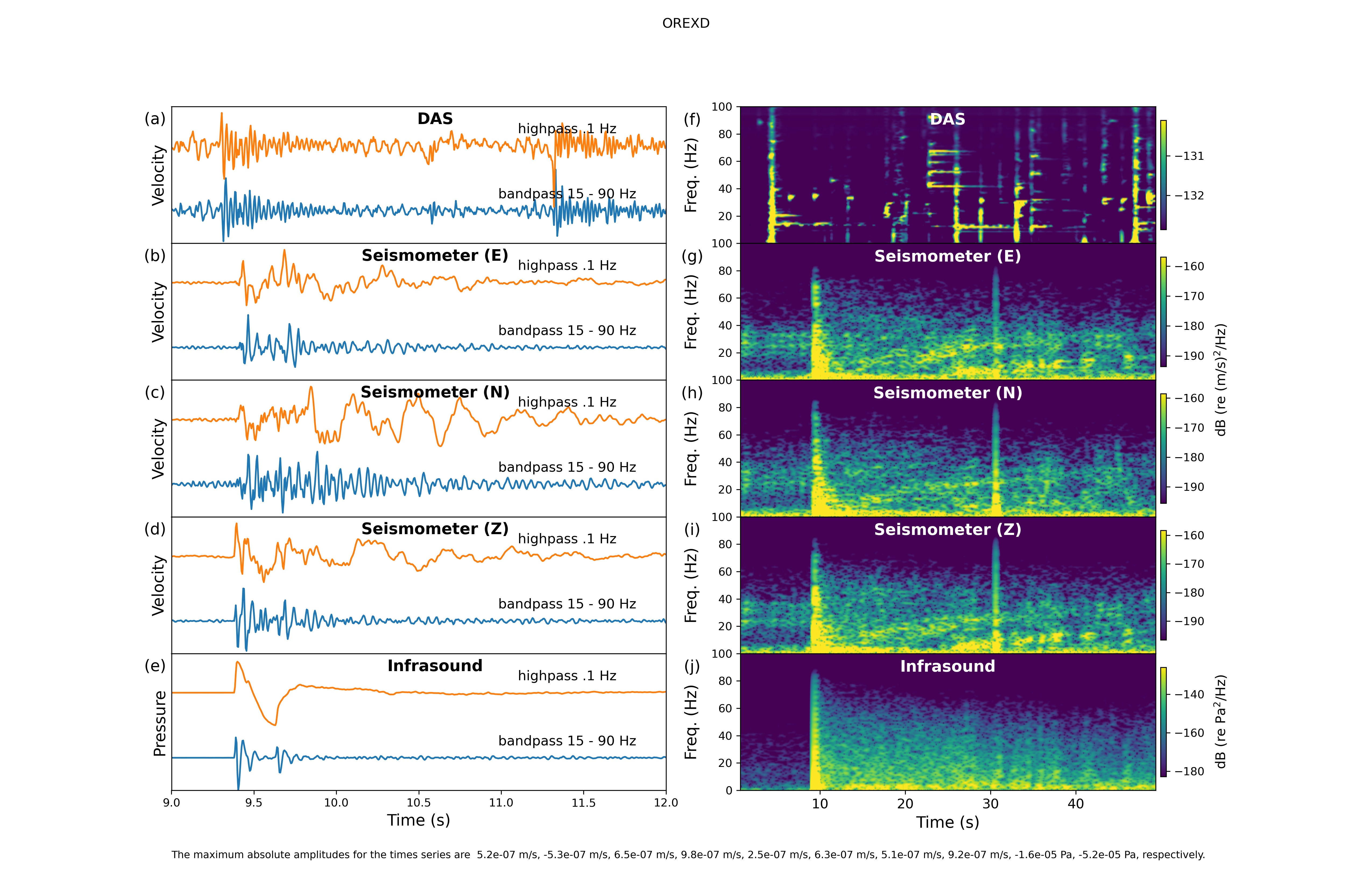}
    \caption{Time series (a--e) and spectrograms (f--j) of data for sensors at or near OREXD. DAS data were selected to be near OREXD and have suitable signal-to-noise (see Figure \ref{SuppNoiseCriterion}) and then stacked to further improve SNR. The maximum absolute amplitudes for the times series of velocity data are (a) 4.9e-07$\,$m/s, 5.2e-07$\,$m/s, (b) 6.5e-07$\,$m/s, 9.8e-07$\,$m/s, (c) 2.5e-07$\,$m/s, 6.3e-07$\,$m/s, and (d) 5.1e-07$\,$m/s, 9.2e-07$\,$m/s.} 
    \label{SuppWaveformsSpecD}
\end{figure*}

\begin{figure*}[htp]
    \centering
    \includegraphics[width=\textwidth,trim=4.5cm 1.2cm 4.5cm 2cm, clip]{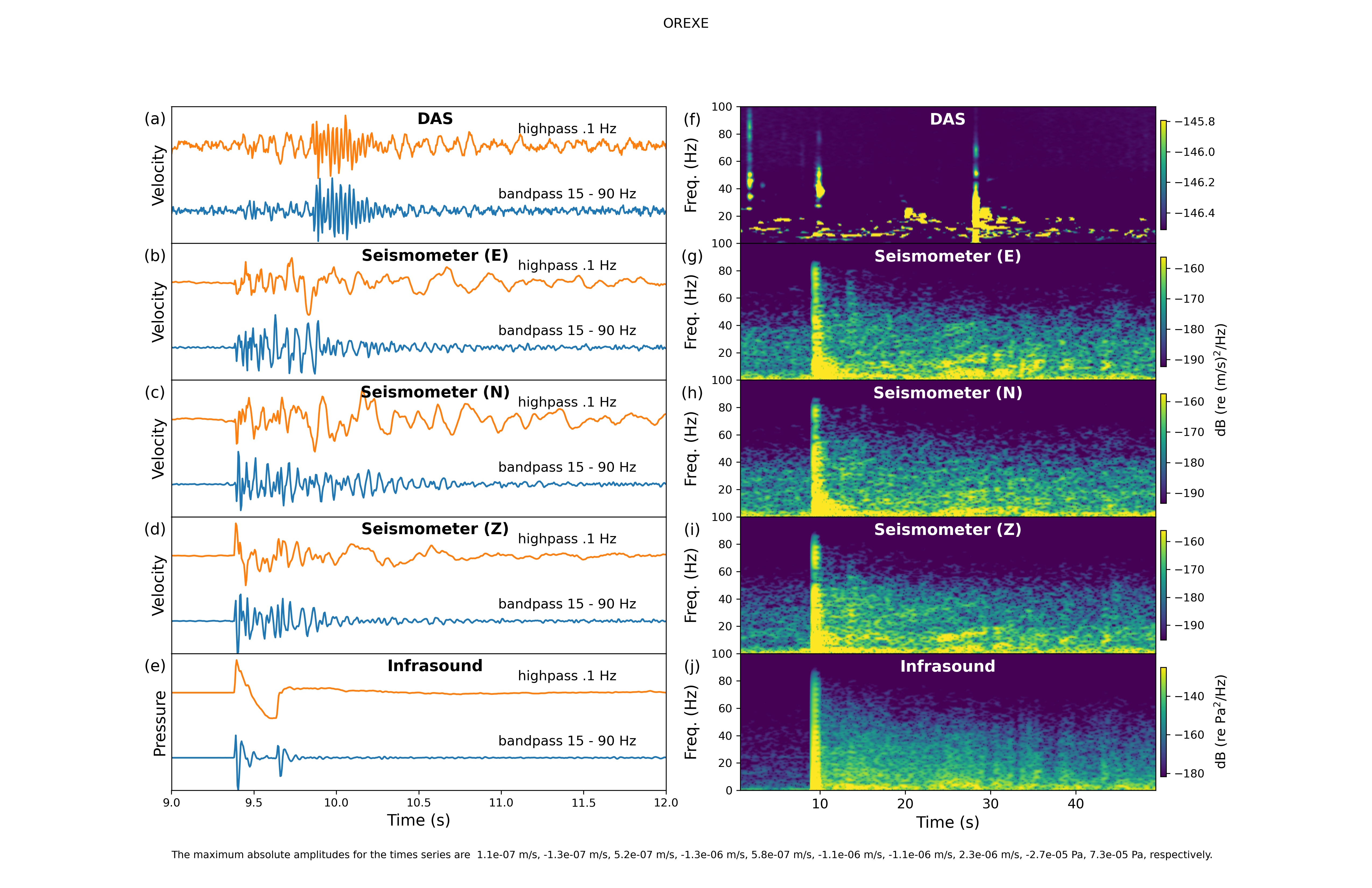}
    \caption{Time series (a--e) and spectrograms (f--j) of data for sensors at or near OREXE. DAS data were selected to be near OREXE and have suitable signal-to-noise (see Figure \ref{SuppNoiseCriterion}) and then stacked to further improve SNR. The maximum absolute amplitudes for the times series of velocity are (a) -1.3e-07$\,$m/s, -1.4e-07$\,$m/s, (b) 5.2e-07$\,$m/s, -1.3e-06$\,$m/s, (c) 5.8e-07$\,$m/s, -1.1e-06$\,$m/s, and (d) -1.1e-06$\,$m/s,  2.3e-06$\,$m/s.}
    \label{SuppWaveformsSpecE}
\end{figure*}

\begin{figure*}[htp]
    \centering
    \includegraphics[width=0.8\textwidth,trim=0cm 0cm 0cm 2cm, clip]{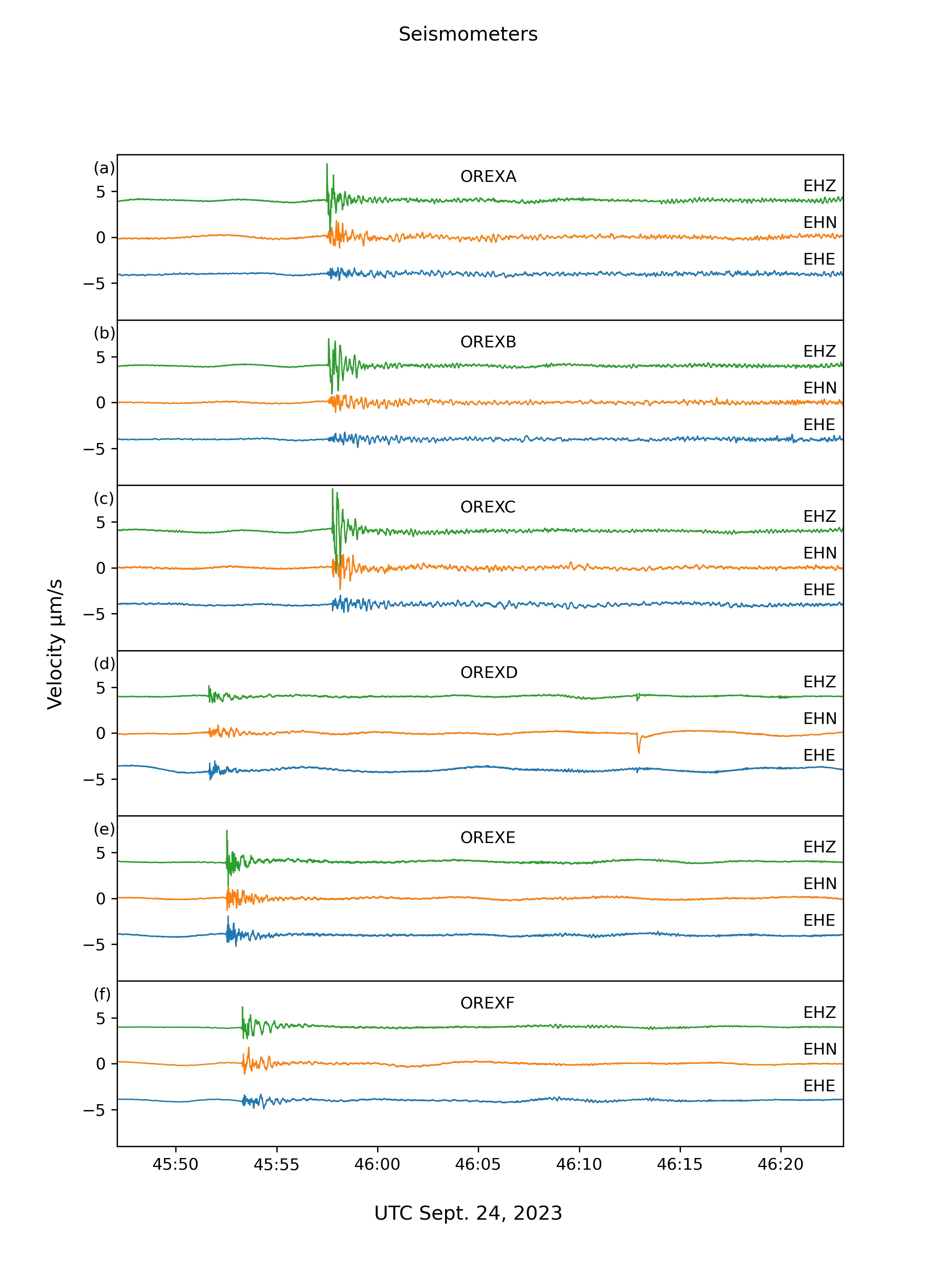}
    \caption{The arrival of the signal generated by the OSIRIS-REx Sample Return Capsule as recorded by six, 3-channel seismometers.}
    \label{SuppAllSeismic}
\end{figure*}

\begin{figure*}[htp]
    \centering
    \includegraphics[width=\textwidth,trim=0.0cm 0cm 0.0cm 1.5cm, clip]{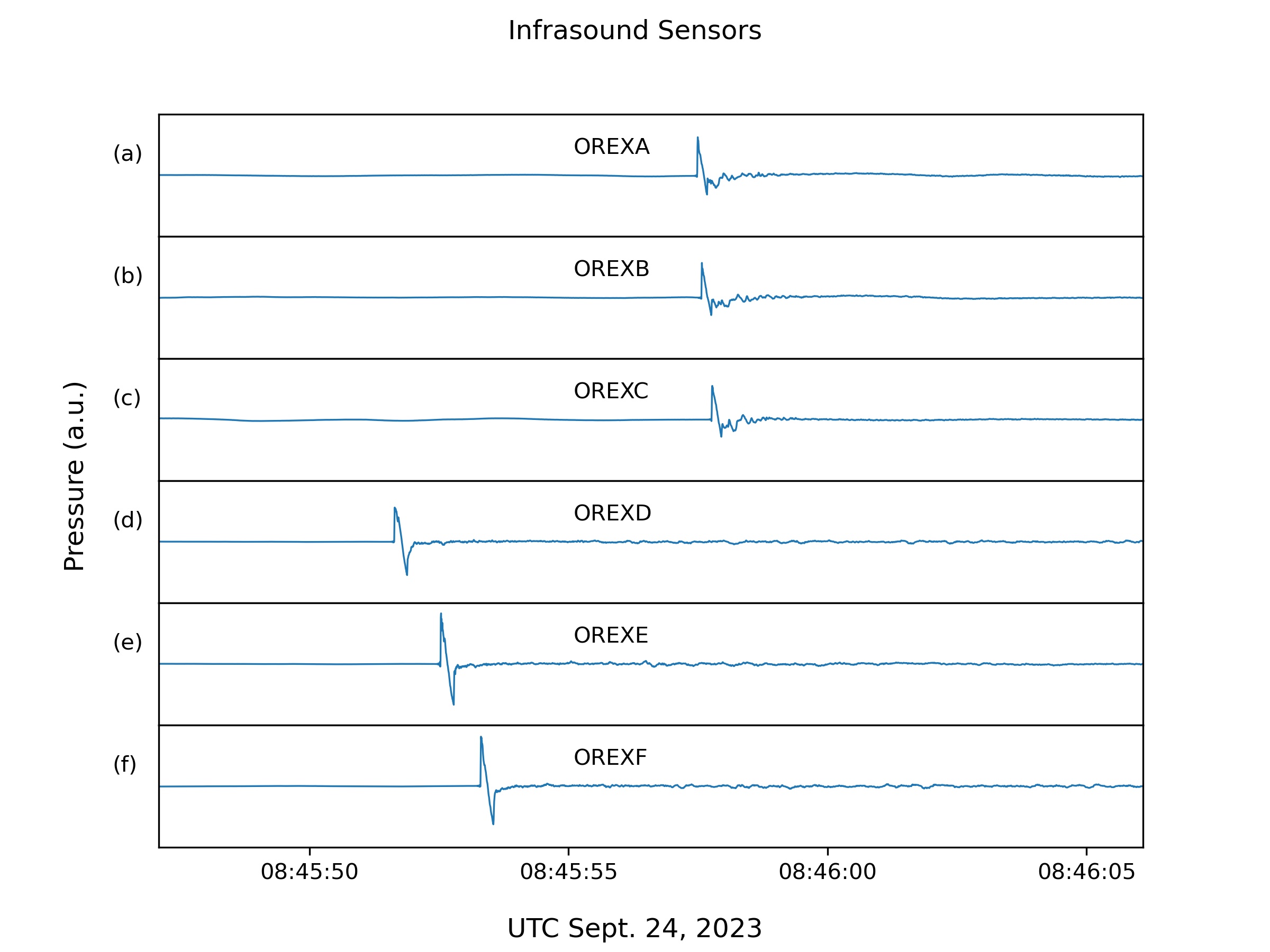}
    \caption{The arrival of the signal generated by the OSIRIS-REx Sample Return Capsule as recorded by six infrasound sensors. Pressure is plotted in arbitrary units (a.u.).}
    \label{SuppAllInfra}
\end{figure*}

\begin{figure*}[htp]
    \centering
    \includegraphics[width=\textwidth,trim=0cm 0cm 0cm 0cm, clip]{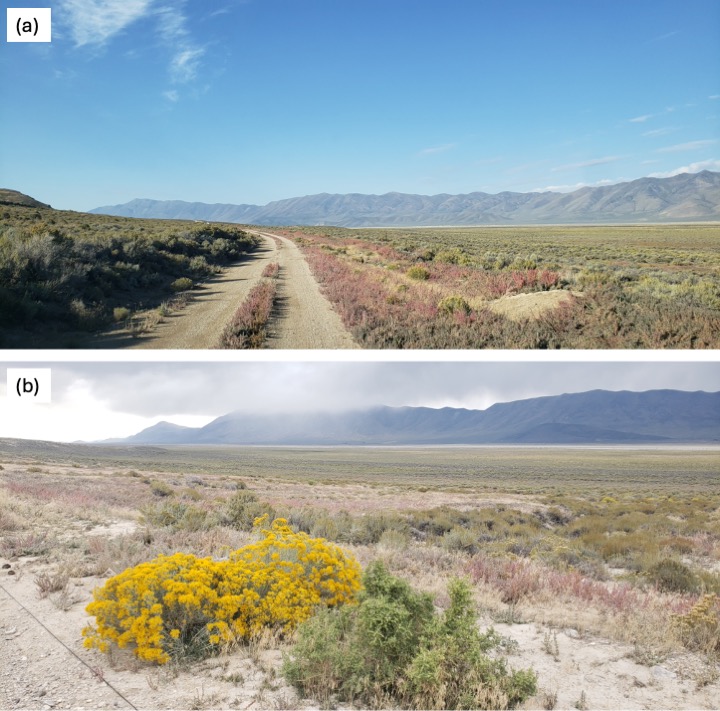}
    \caption{Newark Valley. (a) Dirt road along which the DAS cable was placed. (b) We initially placed the cable beside the dirt road near the vegetation as in this photo. The night before the re-entry, we moved the cable onto one of the tracks in the road (not pictured).}
    \label{SuppDASPlacementPhoto}
\end{figure*}

The supplemental material consists of one table and seven figures. Table \ref{SuppLocationTable} contains location information for the six co-located seismometer/infrasound sensor pairs. Figure \ref{SuppDASCablePhoto} shows the two cable types used in this experiment. Figure \ref{SuppNoiseCriterion} shows which DAS channels in Newark Valley are used for comparison to the nearby seismoacoustic data. In Figures \ref{SuppWaveformsSpecD} and \ref{SuppWaveformsSpecE}, the stacked DAS data are plotted along with the nearby seismic and infrasound data. The data in these plots are processed in the same manner as described in the main text for Figure \ref*{waveformsSpecF}. Figures \ref{SuppAllSeismic} and \ref{SuppAllInfra} show the arrival signal as recorded on all seismometers and infrasound sensors in our study. Figure \ref{SuppDASPlacementPhoto} is an example of placement location of the DAS cable in Newark Valley, where the cable was placed on a dirt road. 

\begin{table}[htp]
\caption{Co-located seismometer and infrasound sensor locations.}
    \centering
    \begin{tabular}{c c c}
    \textbf{Station Name}  & \textbf{Latitude (\textdegree N)} & \textbf{Longitude (\textdegree E)} \\
    \hline
         OREXA & 39.6109883 & -116.002932 \\
         OREXB & 39.60899   & -116.011737 \\
         OREXC & 39.6040433 & -116.004643 \\
         OREXD & 39.7372017 & -115.674093 \\
         OREXE & 39.7043    & -115.676033 \\
         OREXF & 39.6858783 & -115.676975 \\
    \hline
\end{tabular}
 \label{SuppLocationTable}
\end{table}

